\documentstyle[12pt]{article}
\begin{document}
\parindent20pt
\parskip10pt

\hspace*{2in} {Submitted to Phys. Rev. Lett. on May 18, 1998}\\

\vspace{30pt}

\centerline{\bf \Large Bremsstrahlung Spectrum in $\alpha$ Decay }

\vspace{20pt}

\centerline {M.I.Dyakonov *)}
\vspace{20pt}
 \noindent Laboratoire de Physique Math\'ematique cc 070, Universit\'e Montpellier II, 34095\\ Montpellier, France.
E-mail: {\em dyakonov@lpm.univ-montp2.fr}
\vspace{20pt}

Using our previous approach to electromagnetic emission during tunneling,
an explicit, essentially classical, formula describing the bremsstrahlung
spectrum in $\alpha$ decay is derived.  The role of tunneling motion in photon
emission is discussed.  The shape of the spectrum is a universal function
of the ratio $E_{\gamma}/E_{0}$ , where $E_{\gamma}$ is the photon energy and $E_{0}$ is a characteristic energy depending only on the nuclear charge and the energy of the $\alpha$  particle.


\vspace{20pt}

During the $\alpha$ decay of a nucleus the $\alpha$ particle tunnels through the Coulomb barrier and is accelerated beyond the classical turning point to
its final energy.  Thus electromagnetic radiation should be emitted during
the process.  In spite of the fundamental nature of the $\gamma$ emission during$\alpha$ decay, the corresponding bremsstrahlung spectrum was neither measured, nor considered theoretically until recent years.  Normally soft $\gamma$ emission accompanying the Coulomb interaction of heavy particles is well described by classical electrodynamics. The case of $\alpha$ decay is somewhat special,
since part of the "`trajectory"' of the $\alpha$ particle lies within the
underbarrier region and may be described classically only in some limited
sense, as a motion in imaginary time.  While the golden rule of quantum
mechanics gives a straightforward recipe to calculate the emission
probability, intuitively one would like to understand whether tunneling may
be somehow incorporated in the general framework of classical
electrodynamics and whether it makes any sense to say that photons are
emitted during tunneling motion.
 
In Ref. [1] the electromagnetic radiation by a charge tunneling
through a potential barrier was considered under the conditions that the
motion is quasiclassical, i.e. the barrier is smooth compared to the
particle wave length, and that the energy of the emitted photons is small
compared to the particle energy.  It was shown that in this case the
emission spectrum is described by the well known classical formula,
involving the Fourier transform of the particle acceleration, with the only
difference that the Fourier integral should be taken along a special
contour in the complex plane of the time variable. The resulting formula
for the bremsstrahlung spectrum per tunneling particle is classical (does
not contain Planck constant), even though it describes radiation emitted
in the tunneling process, which is unknown to classical mechanics.  The
result of Ref. [1] may be directly applied to $\alpha$ decay.

On the experimental side, D''Arrigo $et$ $al$ [2] were the first to report
bremsstrahlung in $\alpha$ decay.  Their data for $^{226}$Ra and $^{214}$Po gaveemission probabilities much larger than those predicted by the so-called Coulomb
acceleration model, in which only acceleration outside the Coulomb barrier
is taken into account, but close to those predicted by the sudden
acceleration model.  These results are strange since even the Coulomb
acceleration model grossly overestimates the emission probability.
Recently Kasagi $et$ $al$ [3] measured the $\gamma$ emission associated with the $\alpha$ decay of $^{210}$Po, obtaining probabilities which are almost two orders of magnitude smaller than those of Ref. 2, and which have a quite different
dependence on the photon energy.  Kasagi $et$ $al$  applied the general formula
of Ref. [1] and by numerical calculations obtained a good fit to their
experimental data.  However, numerical calculations for the specific case
of $^{210}$Po do not allow to understand the general features of the spectrum
and its dependence on the nuclear charge and $\alpha$ particle energy.  Thus,
there is a need for an explicit theoretical formula describing the
bremsstrahlung spectrum.

Papenbrock and Bertsch [4] considered the problem and arrived to the conclusion that the main contribution to the photon emission stems from Coulomb acceleration, while the contribution of tunneling is negligible, contrary to the point of view expressed in Refs. [1] and [3].  Their numerical results also agree well with the experimental data of Ref. [3].

The purpose of the present Letter is to derive an explicit formula for the bremsstrahl\- ung spectrum in $\alpha$ decay and to elucidate the role of tunneling.  It will be shown that taking tunneling motion into account is very important to obtain the correct description of the bremsstrahlung.

The $\alpha$ particle moves in the Coulomb potential $U(r)=2Ze^2/r$, which is truncated at the nuclear radius $r_{0}$.  The classical turning point is defined by $r_{c}=2Ze^2/E_{\alpha}$, where 2$e$ is the charge of the $\alpha$ particle, $Ze$ is the charge of the daughter nucleus, $E_{\alpha}= mv^2/2$ is the $\alpha$ particle energy, $m$ is the reduced mass, and $v$ is the relative velocity at $r=\infty$.  We assume $r_{0}$ to be much smaller than $r_{c}$.  The initial state is described by an outgoing wave with the asymptotic form $\propto (1/r)$exp$(ik_{i}r)$, with $k_{i}=(2mE_{\alpha})^{1/2}/\hbar$ .  The final state at $r=\infty$   has the form $\;\propto (1/r)$cos$(k_{f}r+\phi)$, $k_{f}=[2m(E_{\alpha}-\hbar \omega)]^{1/2}/\hbar$, $\hbar\omega  =E_{\gamma}$ is the energy of the emitted photon, and $\phi$ is a phase, related to the scattering phase.  In the underbarrier region the wave function of the initial state increases exponentially when $r$ changes from $r_{c}$ inwards, while the wave function of the final state decreases from $r_{c}$ inwards. (For a detailed discussion of the initial and final states entering the matrix element for photon emission see Ref. [4]).

The following conditions will be assumed:
 \begin {equation} k_{i}r_{c} = 2Ze^2/\hbar v >>1,  \end {equation}
 \begin {equation}\hbar\omega << E_{\alpha}.      \end {equation}

Eq. (1) says that the motion is quasiclassical which is generally very well
justified, e.g. for $Z=50, E_{\alpha} =4$ MeV we have $k_{i}r_{c}=100$.  Eq. (2) means that the energy losses are small, and we will see below that the main body of the bremsstrahlung spectrum lies in this energy range.

We can now use the quasiclassical approximation for the initial and final wave functions by replacing $kr$ with $\int k(r)dr$ , where  $k_{i}=[2m(E_{\alpha}-U(r)]^{1/2}/\hbar$  ,  $k_{f}=[2m(E_{\alpha}-\hbar \omega -U(r))]^{1/2}/\hbar$   .  The reflected part of the final state wave function with the asymptotic behavior $\propto$ exp$(-ik_{f}r)$ may be neglected [1], since its contribution to the matrix element is exponentially small, compared to the contribution of the outgoing wave $\propto$ exp$(ik_{f}r)$.  To the first order in the small ratio $\hbar\omega/E_{\alpha}$ we have $k_{i}(r)-k_{f}(r)=\hbar\omega/v(r)$, where $v(r)$ is the classical velocity at point $r$.
Proceeding further as in Ref. 1, one finally obtains the classical formula
[5] describing the bremsstrahlung spectrum:
\begin {equation}\frac {\partial {\cal E}}{\partial \omega}=\frac {2}{3\pi} \frac {(Z_{eff}e)^2}{c^3} |a_{\omega}|^2, \end {equation}
where $\partial {\cal E}/\partial \omega$ is the total energy emitted per decay per unit frequency range, $Z_{eff}=2(A-2Z)/(A+4)$ is the effective charge, $A$ is the mass number of the daughter nucleus,
\begin {equation} a_{\omega}=\int_{C} a(t) \exp (-i\omega t)dt \end {equation}
is the "Fourier transform" of the acceleration $a(t)$, the integral is taken
along the contour $C$ which is defined by the manner in which the time
variable $t=\int_{r_{c}}^{r} dr  /v(r)$  changes when $r$ runs from $0$ to $\infty$ along the real axis.  In the underbarrier region $t$ is imaginary, and it is easy to calculate the value of $t$ for $r=0$: $t=-i\tau$, where $\tau=(\pi/2)r_{c}/v$  is the tunneling time.  This value practically does not depend on the details of the potential curve at small distances and may be calculated by extending the Coulomb potential down to $r=0$.  In the following we will use the relation $a_{\omega} =i\omega v_{\omega}$, where $v_{\omega}$ is defined through the classical velocity $v(t)$ analogous to Eq. (4).

The function $v_{\omega}$ generally depends on the details of the potential
function in the region close to the nuclear radius.  However, these details
are irrelevant for sufficiently low frequencies, such that $\omega t_{1}<<1$, where $t_{1}$ is the characteristic time the particle spends in the vicinity of $r_{0}$.  This time can be estimated as $t_{1} \sim r_{0}/v_{0}$, with $mv_{0}^2  \sim Ze^2/r_{0}$, thus $t_{1} \sim \tau (r_{0}/r_{c})^{3/2}$,
and the condition $\omega t_{1}<<1$ may be rewritten as $\omega\tau<<(r_{c}/r_{0})^{3/2}$.  Since, as shown below, the main part of the bremsstrahlung spectrum lies in the region $\omega\tau \sim 1$, and since $r_{c}/r_{0}>>1$, it can be seen that the structure of the
potential function in the vicinity of the nuclear radius may influence only
the high-energy tail of the spectrum where the emission probability is
already exponentially small.  Thus, the main body of the spectrum may be
calculated by extending the Coulomb potential down to $r=0$.

Introduce the dimensionless coordinate $\xi=r/r_{c}$, the dimensionless time $\eta=vt/r_{c}$, and the dimensionless frequency $\nu =(1/2)\omega  r_{c} /v= \omega Ze^2/(E_{\alpha}v)$. Then $\partial {\cal E}/\partial \omega$  may be expressed as
                         
\begin {equation} \frac {\partial {\cal E}}{\partial \omega}=\frac {2(Z_{eff}e)^2 v^2}{3\pi c^3} |J(\nu)|^2, \end {equation}                               

 \begin {equation} J(\nu)=2i\nu \int_{-i\pi /2}^{\infty} d\eta (d\xi /d\eta) \exp (-2i\nu \eta) \end {equation}
  
\noindent where the function $\xi (\eta)$ is the solution of the (dimensionless)Newton equation $d^2\xi /d\eta^2=(2\xi ^2)^{-1}$ with the conditions $\xi (0)=1, (d\xi /d\eta)_{\eta=\infty}=1$. 
 
This may be compared with the results for two related classical problems: a) $Coulomb$ $acceleration$ $model$ in which only acceleration outside the Coulomb barrier is taken into account.  The lower limit for the integral in Eq. (6) should be $\eta=0$, instead of $\eta=-i\pi/2$.  b) $Head-on$ $collision$ $of$ $two$ $particles$ $with$ $charges$ $2e$ $and$ $Ze$.  In this case the lower limit in Eq. (6) should be $\eta=-\infty$, instead of $\eta=-i\pi/2$.

Introducing a new variable $z$, such that $\xi=(1+$cosh$z)/2,  \eta =(z+$sinh$z)/2$, we obtain

\begin {equation}J(\nu)=i\nu \int_{-i\pi}^{\infty} dz \sinh z \exp [-i\nu(z+\sinh z)]. \end {equation}

The integration contour in Eq. (7) goes from $-i\pi$ to $0$ along the imaginary
axis and from $0$ to $\infty$ along the real axis.  It may be safely deformed to go from $-i\pi$ to $-i\pi+\infty$ (see Ref. 5).  Finally, introducing $x=z+i\pi$, we get

\begin {equation} J(\nu)=-i\nu \exp (-\pi\nu) \int_{0}^{\infty} dx \sinh x \exp [i\nu(\sinh x-x)]. \end {equation}

By integration by parts $J(\nu)$ may be also expressed in a form which is more suitable for numerical calculations:
 \[J(\nu)=\exp (-\pi\nu )[1-i\nu I(\nu)],\] 
\noindent where
 \begin {equation} I(\nu) = \int_{0}^{\infty} dx (1-e^{-x}) \exp [i\nu (\sinh x-x)]. \end {equation}

\noindent The integral in Eq. (9) converges more rapidly, than the one in Eq. (8). 

Eqs. (5) and (8) give an explicit formula for the bremsstrahlung spectrum in $\alpha$ decay, provided the conditions given by Eqs. (1) and (2) are fulfilled, and this is the main result of this work.  For the head-on collision problem, mentioned above, $J(\nu)$ should be replaced by $J_{coll}(\nu)$, where $J_{coll}(\nu)$ is given by the same Eq. (8) but with the lower limit of integration replaced by $-\infty$ (see Ref. [5]).  Thus $J_{coll}(\nu)=2$Re$J(\nu)$.

The Coulomb acceleration model, which neglects tunneling, predicts a spectrum that falls down at high frequencies as $\omega ^{-2}$.  The reason is that in this model the acceleration has a singular behavior, increasing step-wise from a zero value at $t<0$ to a finite value at $t=0$.  Taking tunneling into account makes the function $a(t)$ smooth, and as a result, the spectrum falls down exponentially.  One might say that there is a destructive interference between electromagnetic fields emitted during the tunneling motion at $r<r_{c}$ and during the classical acceleration stage at $r>r_{c}$.  (Similarly, in the classical Coulomb collision problem, the interference between the incoming and outgoing parts of the trajectory also produces an exponentially decreasing spectrum at high frequencies [5]). Thus, as can be also seen from Eqs. (4), (6), taking account of the
tunneling motion is very important to obtain the correct description of
bremsstrahlung.
  
The opposite conclusion reached by Papenbrock and Bertsch [4] is due to a misinterpretation.  When applied to the quasiclassical limit of the acceleration matrix element (Eq. (8) above), their reasoning goes as follows.  Noticing that Im$J(\nu)=0$ at $\nu=0$ [6], they make the statement that Im$J(\nu)$ is {\em numerically} small compared to Re$J(\nu)$ for arbitrary $\nu$ (in fact, for $\nu>>1$, the ratio Im$J(\nu)/$Re$J(\nu)$ is equal to $-1/\sqrt {3}$ ).  On these grounds the imaginary part of $J(\nu)$ is discarded, while the real part, as shown above,
can be expressed as $(1/2)J_{coll}(\nu)$.  Since calculation of $J_{coll}(\nu)$ does not involve tunneling motion, the conclusion is made, that the contribution of tunneling is negligible.  In fact, neglecting Im$J(\nu)$ is equivalent to
replacing the true initial wave function $\propto \; $exp$(ik_{i}r)$ at $r=\infty$) by its real part which describes an $\alpha$ particle scattered by the nucleus ($\propto \;$cos$(k_{i}r)$ at $r=\infty$), but normalized to 1/2 incoming flux. Not surprisingly, the spectrum for $\alpha$ decay, obtained in Ref. 4 after neglecting Im$J(\nu)$, is equal to 1/4 of the spectrum for the head-on Coulomb collision problem [7].  However, this result has a correct behavior both at low and at high frequencies, the difference with the true spectrum being not very large (25\% for $\nu>>1$), especially when plotted on a logarithmic scale.  Clearly, the fact that the spectrum for the collision problem (with a factor 1/4) does not differ very much from the true spectrum in $\alpha$ decay, does not mean that tunneling is not important.

Papenbrock and Bertsch [4] have also performed a numerical calculation of the exact matrix element, without neglecting its real part and without using the quasiclassical approximation, for the specific case of  $^{210}$Po. The resulting spectrum is very close to the spectrum given by the explicit Eqs. (5), (8), (9) of the present work (see below).  Thus the numerical results of Ref. 4 are correct and are in agreement with the above formulae, which explicitly take into account the tunneling part of the trajectory. However the physical interpretation of these results in Ref. 4, leading to the conclusion that the main contribution to the photon emission stems from the Coulomb acceleration, is erroneous.  In fact, it is the contribution of the tunneling region that accounts for the very large difference between the true spectrum and the predictions of the Coulomb acceleration model.

It is easy to find the asymptotic form of $J(\nu)$ both for small and for large values of $\nu$.  For $\nu<<1 \hspace*{0.1in} J(\nu)=1-\pi \nu/2-i\nu$ln$(2\gamma/\nu)$, where $\gamma$ =1.781, so that $|J(\nu)|^2 \approx 1-\pi\nu$, while for $\nu>>1$

\begin {equation} J(\nu)=-i \exp (i\pi/3)\Gamma (2/3)(4\nu /3)^{1/3} \exp (-\pi \nu) \end {equation}

\begin {equation} |J(\nu)|^2 = \Gamma^2 (2/3)(4\nu /3)^{2/3} \exp (-2\pi \nu )=2.22 \nu ^{2/3} \exp (-2\pi \nu). \end {equation}

\noindent It follows from Eqs. (10), (11) that for $\nu>>1 \hspace*{0.1in} |J_{coll}(\nu)|^2=3|J(\nu)|^2$.  Fig. 1 shows the normalized spectrum, given by the function $|J(\nu)|^2$, calculated numerically, together with the asymptotic curves for $\nu<<1$ and $\nu>>1$.  It can be seen that Eq. (11) becomes a very good approximation for $\nu>0.5$.  In this region the $|J(\nu)|^2$ dependence is dominated by the exponential factor in Eq. (11), which may be expressed as exp$(-E_{\gamma} /E_{0})$), where
\[E_{0}=\frac {E_{\alpha}}{2\pi} \frac {\hbar v}{Ze^2} \]
is a characteristic energy, $2\pi\nu=E_{\gamma} /E_{0}$. For the experimental conditions of Ref. [3] ($Z=82, E=5.3$ MeV) one finds $E_{0} \approx 75$ keV.  Note that the shape of the spectrum is a universal function of $E_{\gamma} /E_{0}$.

The experimentally measured quantity is the differential probability of photon emission per decay $\partial ^2 N/\partial E_{\gamma} \partial \Omega =(4\pi\hbar E_{\gamma})^{-1} \partial {\cal E}/\partial \omega$. Thus

\begin {equation} \frac {\partial ^2 N}{\partial E_{\gamma} \partial \Omega} = \frac {Z_{eff}^2}{3\pi^2} \frac {e^2}{\hbar c} \frac {E_{\alpha}}{mc^2} \frac {1}{E_{\gamma}} |J(\nu)|^2. \end {equation}

\noindent This quantity is plotted in Fig. 2 as a function of $E_{\gamma}$ for the case of $\alpha$ decay of $^{210}$Po (heavy line), together with the experimental data taken from Ref. [3].  The result of the numerical calculation in Ref. [3] is presented by the dotted line.  This calculation was done by using the general formula of Ref. [1], and thus should give the same result as the present work.
However, one can see that the two results are qualitatively different:
a)the slopes at low $\gamma$ ray energies differ by a factor $\sim 2$ and b)while the present theory gives a spectrum that falls  down exponentially at high $E_{\gamma}$ , the calculation of Ref. [3] gives a broad maximum around $E_{\gamma} \sim 500$ keV.  The difference in the slopes can not be easily understood.  As to the high energy part of the spectrum, we should note that there is a limitation of the present theory related to our neglecting the details of the inner
barrier region and the extension of the Coulomb potential down to $r=0$.  As
explained above, this is justified if $E_{\gamma}<<E_{0}(r_{c}/r_{0})^{3/2} \approx 860$ keV ($r_{c}/r_{0} \approx 5$ for $^{210}$Po).  For energies comparable to or higher than this value the shape of the spectrum should depend on the parameters of the inner barrier region.  Thus, the maximum on the curve calculated in Ref. [3] (where the cutoff of the Coulomb potential at $r=r_{0}$ was taken into account) might be related to this circumstance.  More precise experimental data are needed to determine unambiguously the true slope at low energies, as well as the existence of any specific features at the high energy tail of the spectrum.
  
In summary, an explicit, essentially classical, formula, describing the bremsstrahlung spectrum in a decay, was derived, and it was shown that taking the tunneling motion into account is crucial for the correct description of the spectrum.  The shape of the spectrum is a universal function of the ratio $E_{\gamma} /E_{0}$ , where $E_{\gamma}$  is the photon energy and a $E_{0}$ is a characteristic energy depending only on the nuclear charge and the energy of the $\alpha$  particle.

I thank Michael Shur for his hospitality during my stay at Rensselaer Polytechnic Institute, where this work was completed.  I appreciate useful discussions with Thomas Papenbrock.

\newpage

{\bf {References}}

\vspace {20pt}
\noindent *) On leave from A.F. Ioffe Physico-Technical Institute, St. Petersburg, Russia\\
$[1]$ M.I. Dyakonov and I.V. Gornyi, Phys. Rev. Lett. 76, 3542 (1996)\\
$[2]$ A. D''Arrigo et al, Phys. Lett. B 332, 25 (1994)\\
$[3]$ J. Kasagi et al., Phys. Rev. Lett. 79, 371 (1997)\\
$[4]$ T. Papenbrock and G.F.Bertsch. Phys. Rev. Lett. 80, 4141 (1998)\\
$[5]$ L.D. Landau and E.M. Lifshitz, The classical theory of fields (Pergamon
Press, Oxford, 1975), p. 181\\
$[6]$ Indeed, tunneling does not contribute to the bremsstrahlung spectrum at
very low frequencies ($\nu<<1$), because, as can be seen from Eq. (4), at $\omega=0$ the contribution of the tunneling region to aw is equal to the difference
of the initial relative velocity and the velocity at the barrier exit point
$r_{c}$ , both of which are equal to zero.  This is no more true for frequencies
$\nu$ on the order of, or much greater than 1.  Note that the real part of the
acceleration matrix element in Ref. [4] corresponds to the imaginary part
of $J(\nu)$, and vice versa.\\
$[7]$ Ref. [4] does not contain an explicit formula for the spectrum in the
quasiclassical approximation.  It may be derived by taking the appropriate
limit in Eq. (14) of Ref. [4] and by correcting the exponent in this
equation by a factor of 2.

\vspace {30pt} 

{\bf {Figure captions}}

\vspace {20pt}

Fig. 1.  Normalized bremsstrahlung spectrum. ({\em a}) - normal scale: {\em 1} -
numerical calculation, {\em 2} - low frequency asymptote $|J(\nu)|^2=1-\pi \nu$, {\em 3} - high frequency asymptote, Eq. (11). ({\em b}) - logarithmic scale: {\em 1} - numerical calculation, {\em 3} - high frequency asymptote, Eq. (11), {\em 4} - result for Coulomb acceleration model, {\em 5} - high frequency asymptote for Coulomb acceleration model $|J(\nu)|^{2}=(4\nu)^{-2}$.\\

\vspace {10pt}

Fig. 2. $\gamma$ ray emission probability in $\alpha$ decay of $^{210}$Po.  Results of present theory (heavy solid line) are presented together with the experimental data points and results of numerical calculation from Ref. [3].  Thin solid line is the high frequency asymptote given by Eqs. (11), (12).

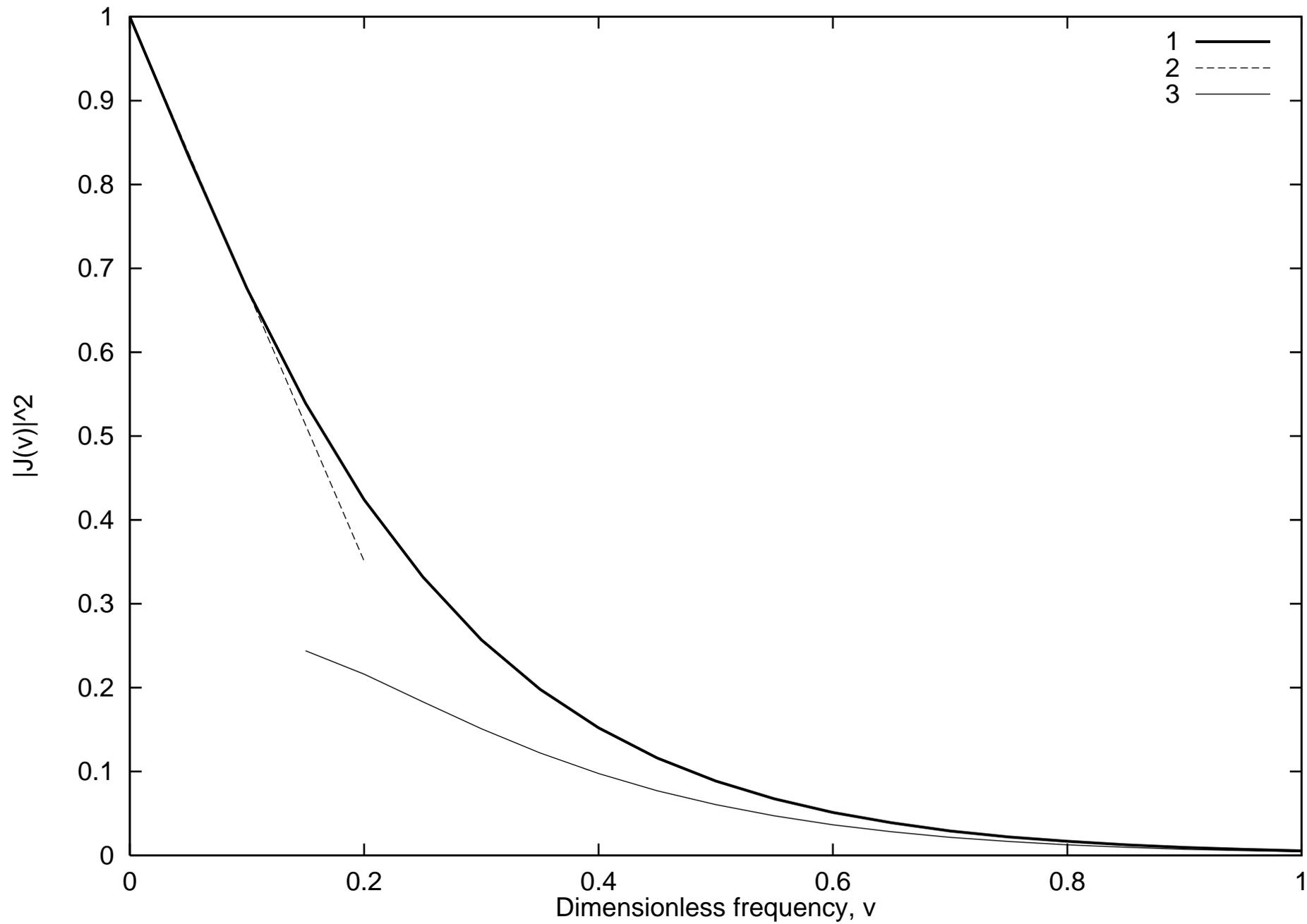
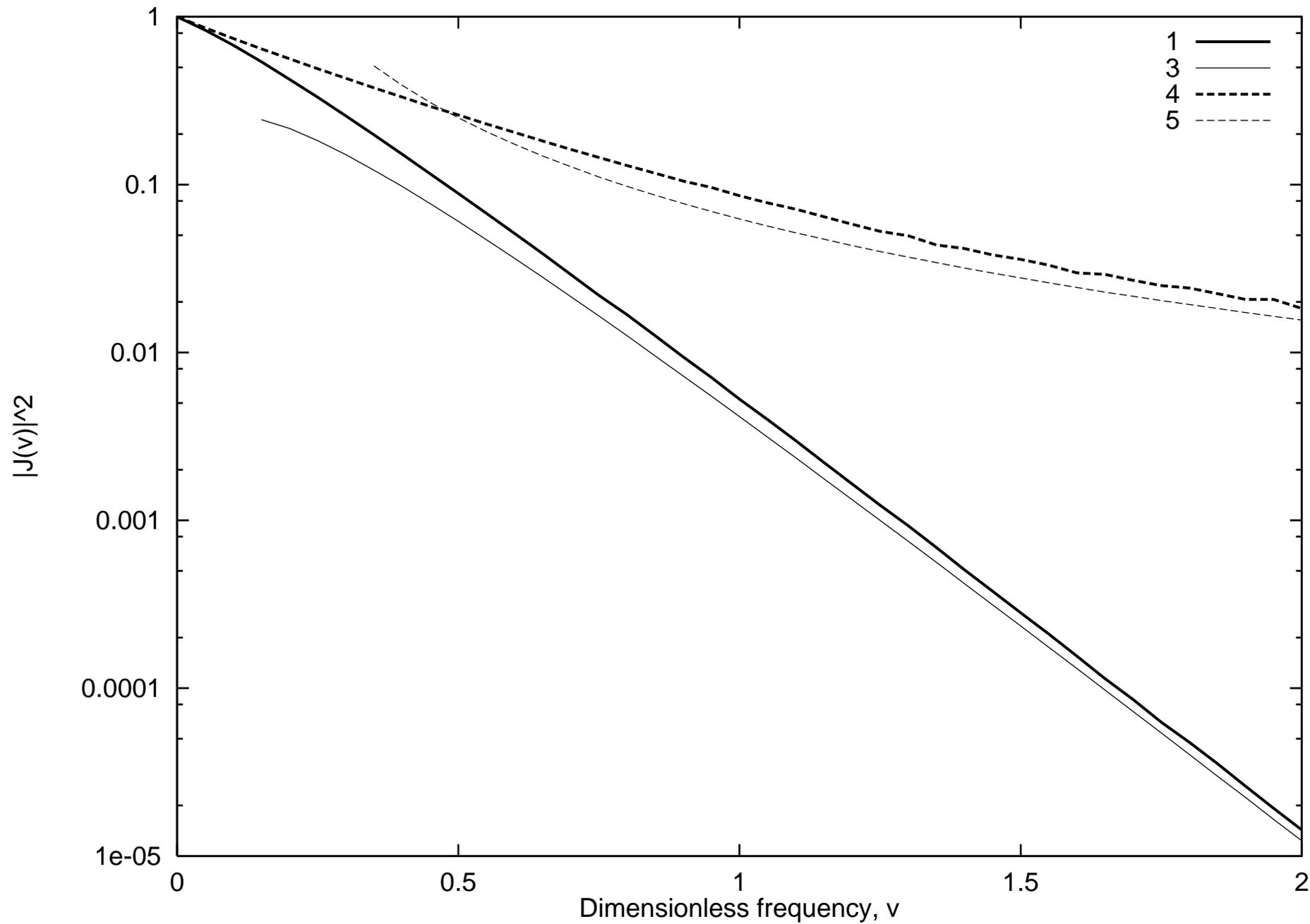
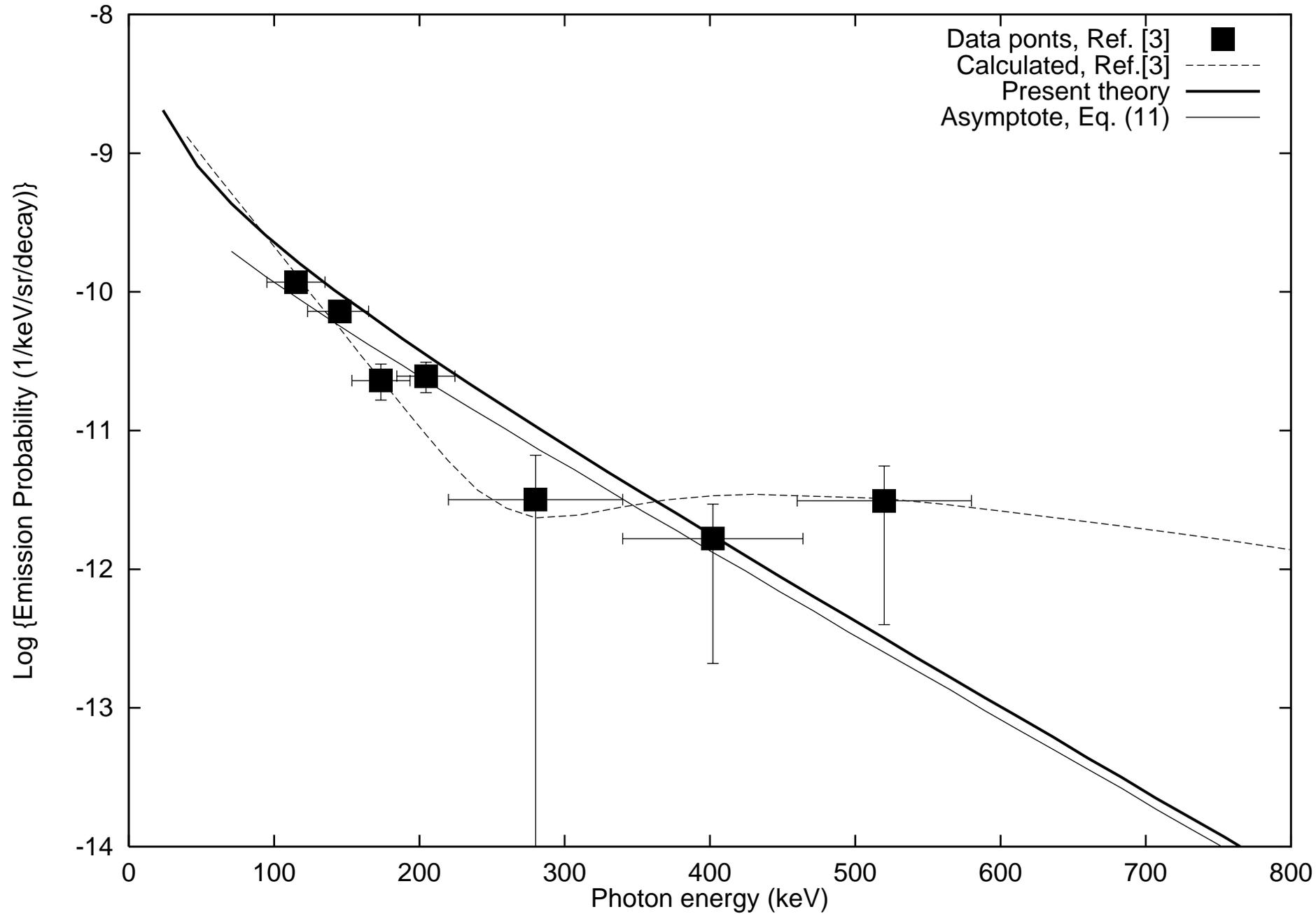
\end{document}